\documentclass{aa}
\usepackage{graphics,epsfig}
\usepackage{txfonts}
\usepackage{xcolor}

\def\simless{\mathbin{\lower 3pt\hbox
     {$\rlap{\raise 5pt\hbox{$\char'074$}}\mathchar"7218$}}}   
\def\simmore{\mathbin{\lower 3pt\hbox
     {$\rlap{\raise 5pt\hbox{$\char'076$}}\mathchar"7218$}}}   

\begin{document}

\title{A jet model for Galactic black-hole X-ray sources: the cutoff
energy--phase-lag correlation}

\subtitle{}

\author
{P. Reig\inst{1,2}, \and N. D. Kylafis\inst{2,1}
}

\institute{
IESL, Foundation for Research and Technology-Hellas, 711 10,
Heraklion,
Crete, Greece
\and 
University of Crete, Physics Department \& Institute of
Theoretical \& Computational Physics, 70013 Heraklion, Crete, Greece
}

\authorrunning{Reig \& Kylafis}
\titlerunning{A jet model for Galactic black-hole X-ray sources}

\offprints{pau@physics.uoc.gr}

\date{Received: \\
Accepted: \\}

  \abstract
{Galactic black-hole X-ray binaries  emit a compact, optically thick,
mildy relativistic radio jet when they are in the hard and hard-intermediate
states, i.e., typically at the beginning and the end of an X-ray outburst. In a
series of papers, we have developed a jet model and have shown, through Monte
Carlo simulations, that our model can explain many observational results.}
{In this work, we investigate  one more constraining relationship
between the cutoff energy and the phase lag during 
the early stages of an X-ray 
outburst of the black-hole X-ray binary GX 339--4: the cutoff 
energy decreases while the phase lag increases during the brightening of
the hard state.}
{We have performed Monte Carlo simulations of Compton upscattering of soft, 
accretion-disk photons in the jet and computed the phase lag between soft
and hard photons and the cutoff energy of the resulting high-energy power law.}
{We demonstrate that our  jet model naturally explains 
the above correlation, with a minor modification consisting of introducing
an acceleration zone at the base of the jet.}
{The observed correlation between the cutoff energy and the phase
lag in the black-hole binary GX 339--4 suggests that 
the lags are produced by the hard component. Here we show
that this correlation arises naturally if Comptonization in the jet
produces these two quantities.}


\maketitle

\section{Introduction} 

The continuum X-ray spectra of black-hole binaries (BHB) are well described by
just two components: i) a soft component, normally modeled as a 
multi-color black-body  component dominating the spectrum below $\sim$10
keV \citep{mitsuda84,merloni00} and whose origin is attributed to a 
geometrically thin, optically thick disk \citep{shakura73} and ii) a
power-law hard tail with an exponential cutoff, which is believed to be the
result of Comptonization of low-energy photons from the disk by energetic
electrons in a configuration that is still under debate. The Comptonizing
medium could be an optically thin, very hot "corona" in the vicinity of the
compact object \citep{titarchuk80,hua95,zdziarski98}, an
advection-dominated accretion flow \citep{narayan94,esin97}, a low angular
momentum accretion flow \citep{ghosh11,garain12}, or the base of a radio jet
\citep{band86,georganopoulos02}. On top of these components, a discrete line
at 6.4 keV is generally observed and attributed to reflection of the hard
X-rays from the accretion disk \citep{fabian89}.


Based on the relative contribution of these two spectral components and the
different shapes and characteristic frequencies of the noise components
in the power spectra, including quasi-periodic oscillations (QPO), 
black-hole systems can be found in several states
\citep{mclintock06,belloni10}, of which the two main ones are called soft
and hard states \citep{done07}. In the soft state, the thermal
blackbody component dominates the energy spectrum with no or very weak
power-law emission \citep{remillard06,dexter12}. Weak power-law
noise  (rms $<$ 5\%) is detected in the power spectrum and sometimes a QPO
at 10--20 Hz \citep{vanderklis06}. In the hard state, the soft
component is weak or absent, whereas the hard tail extends to a few hundred
keV in the form of a power law with photon-number index in the range 1.5 -
2. The power law falls exponentially at a few tens to about a hundred keV
\citep{mclintock06,castro14}. The power spectrum shows strong
band-limited noise with a typical strength of 20\% - 40\% rms and a break
frequency below 1 Hz \citep{homan05,belloni14}. In addition, the
existence of phase (time) lag between the light curves at two different
energy bands obtained simultaneously is well established. The magnitude of
this lag strongly depends on Fourier frequency and on the energy bands
considered \citep{miyamoto88,vaughan97,nowak99a,poutanen01,pottschmidt03}.



The hard and hard-intermediate  states are prominent because it is in these
states that a  compact, optically thick, mildy relativistic jet is detected
in the radio band \citep{fender09}, hence the relation between accretion
and outflow in accreting black holes can be best studied.   For the
formation and destruction of jets in black-hole and neutron star binaries,
the reader is referred to \citet{kylafis12}. For an interpretation of the
observed phenomenology of black-hole X-ray transients the reader is
referred to \citet{kylafis15}.

In a series of papers, we have developed a model for the hard state of BHB. We
have shown that Compton upscattering in the jet of soft photons from the
accretion disk can explain a number of observational relations between the
spectral and timing parameters. Our results demonstrate that jets 
play a central role in all the observed phenomena, not only in the radio
emission.

In \citet{reig03} (hereafter Paper I), we reproduced the X-ray energy
spectra and the dependence of time lag on Fourier frequency and
investigated how the optical depth and extent of the base of the jet affect
the spectral continuum. For simplicity, we assumed that the jet has a
constant flow velocity $v_\parallel$ and in the rest frame of the flow the
electrons are mono-energetic, with a velocity component perpendicular to the
magnetic field $v_\perp$.  The observed  parabolic shape of the jet implies
that the density of the electrons  in the jet drops inversely proportional
to the vertical distance $z$ from the black hole. 

\citet{giannios04} (hereafter Paper II)
showed that both the hardening of
the high-frequency power spectra and, equivalently,
the narrowing of the auto-correlation
function with photon energy observed in Cygnus X-1 can be explained 
by simply assuming that the electrons close to the core of the jet are more
energetic than those at its periphery.  Specifically, $v_\perp$ was assumed
to drop linearly with polar distance from the axis of the jet.

In order to explain the entire spectrum from radio to X-rays,
\citet{giannios05} (hereafter Paper III) assumed that the electrons have
a power-law energy distribution.  This assumption helps explain the radio
part of the spectrum, but has no effect on the Comptonization.  This is 
because the distribution is steep and Comptonization is practically 
performed by the electrons with the lowest Lorentz factor 
$\gamma_{\rm min}$.

In \citet{kylafis08} (hereafter Paper IV),  the correlation 
observed in Cyg X-1 between the
photon index $\Gamma$ and the average time lag was
explained without any additional modification to the model.  Similarly
for the correlation between $\Gamma$ and the characteristic frequencies 
of the Lorenztian peaks \citep{pottschmidt03}.

In this work, we aim at reproducing the correlation between the cutoff
energy $E_c$ and  the phase lag $\phi_{\rm lag}$ between hard and soft
photons, as measured for the BHB GX 339--4 with no or minimal modifications
to our jet model. \citet{motta09} studied the evolution of the high-energy
cut-off in the X-ray spectrum of GX 339--4 across a hard-to-soft transition
and found that the cutoff energy decreased monotonically from 120 to 60 keV
during the brightening of the hard state. \citet{altamirano15} studied the
evolution of the phase lag of GX 339--4 as a function of the position
of  the source in the hardness-intensity diagram ($q$-diagram) and found
that the phase lag increases as the sources moves up in the $q$-diagram
through the hard state. They also showed that $E_c$ and $\phi_{\rm
lag}$ appear to be correlated.

In \S\ 2 we present our model briefly, in \S\ 3 we give the results of our
calculations, in \S\ 4 we discuss our results, and in \S\ 5 we draw 
our conclusions.

\section{The model}

One of the models that we used  in this work (model 1) is a simplified
version of that  used in Paper III.  Since we are not interested in
reproducing the radio spectrum of GX 339-4, we have assumed mono-energetic
electrons in the jet with  Lorentz factor equal to the smallest in the
distribution, namely

\begin{equation}
\label{lorenzt}
\gamma_{\rm min}=\frac{1}{\sqrt{1-(v_{\parallel}^2+v_{\perp}^2)/c^2}},
\end{equation}
where $v_\parallel = v_0 =$ constant 
is the flow velocity of the jet and $v_\perp$ is the 
smallest peprendicular velocity of the electrons in the rest frame of
the flow.
As such, our model is identical to that used in Paper IV.

Simplicity is a preferred quality in models, but there are limits.  
In our model 1 above, we have assumed that the flow velocity in the jet is 
$v_0=$ constant throughout the jet, 
or equivalently that the acceleration region of the jet is infinitesimal.  
This is an unphysical description of the base of the jet.  
For this reason, we have considered a second model (model 2), where
the flow velocity in the jet is given by
\begin{equation}
\label{accel}
v_{\parallel}(z) =
\begin{cases}
(z/z_1)^p ~ v_0  & \text{if } 0< z \leqslant z_1\\
v_0                   & \text{if } z> z_1,
\end{cases}
\end{equation}
where $z_1$ and $p$ are parameters.  In other words, the jet has an
acceleration region of thickness $z_1$ beyond which the flow has 
constant velocity $v_0$, equal to that of model 1.

For a parabolic jet, i.e. one whose radius at height $z$ is 
$R(z)=R_0(z/z_0)^{1/2}$, the electron density in model 1
is inversely proportional to $z$, namely $n_e(z)=n_0(z_0/z)$
(see Paper IV), while
in model 2 it is obtained from the continuity equation.

The fixed parameters of our models and their reference values are:  
the radius $R_0 = 100 r_g$ of the base of the jet, where $r_g=GM/c^2$
is the gravitational radius of the black hole, 
the distance $z_0 = 5 r_g$ of the bottom of the jet from the black hole,
the height $H=10^5 r_g$ of the jet,
the velocity $v_0= 0.8 c$ of the jet, 
the thickness $z_1= 50 r_g$ of the acceleration zone,
the exponent $p=0$ (model 1) or $p=1/2$ (models 2), and the temperature 
$T_{bb} = 0.2$ keV of the soft-photon input.

The parameters of our models that we have varied are:
either the Thomson optical depth along the axis of the jet
$1 \le \tau_\parallel \le 10$ or
the minimum Lorentz factor $2 \le \gamma_{\rm min} \le 2.4$
or both with a linear relation between the two.  Since $v_0$ is 
a constant in our models,
the variation of $\gamma_{\rm min}$ comes from the variation of
$0.35 c \le v_\perp \le 0.42 c$.
  
As the source moves from the hard state to the hard-intermediate one, the
jet weakens \citep{homan05} and eventually disappears at the jet line.  We
interpret this weakening of the jet as a decrease  of the parameter
$\tau_\parallel$.  At the same time, the luminosity increases, cooling of
the jet is enhanced and we interpret this as a decrease of the parameter
$\gamma_{\rm min}$. Therefore $\tau_\parallel$ and $\gamma_{\rm min}$
decrease with time during the initial rise of the outburst.

Since the jet is mildly relativistic, the results also
depend on the angle $\theta$ of observation with respect to the jet axis.
As in Papers II--IV, we will focus in an intermediate range of observing
angles $0.2 < \cos \theta < 0.6$. Practically, for the Monte Carlo
simulation this means that we count only photons that leave the jet in this
range of angles. 
 \begin{figure}[!t]
   \centering
   \includegraphics[width=8cm]{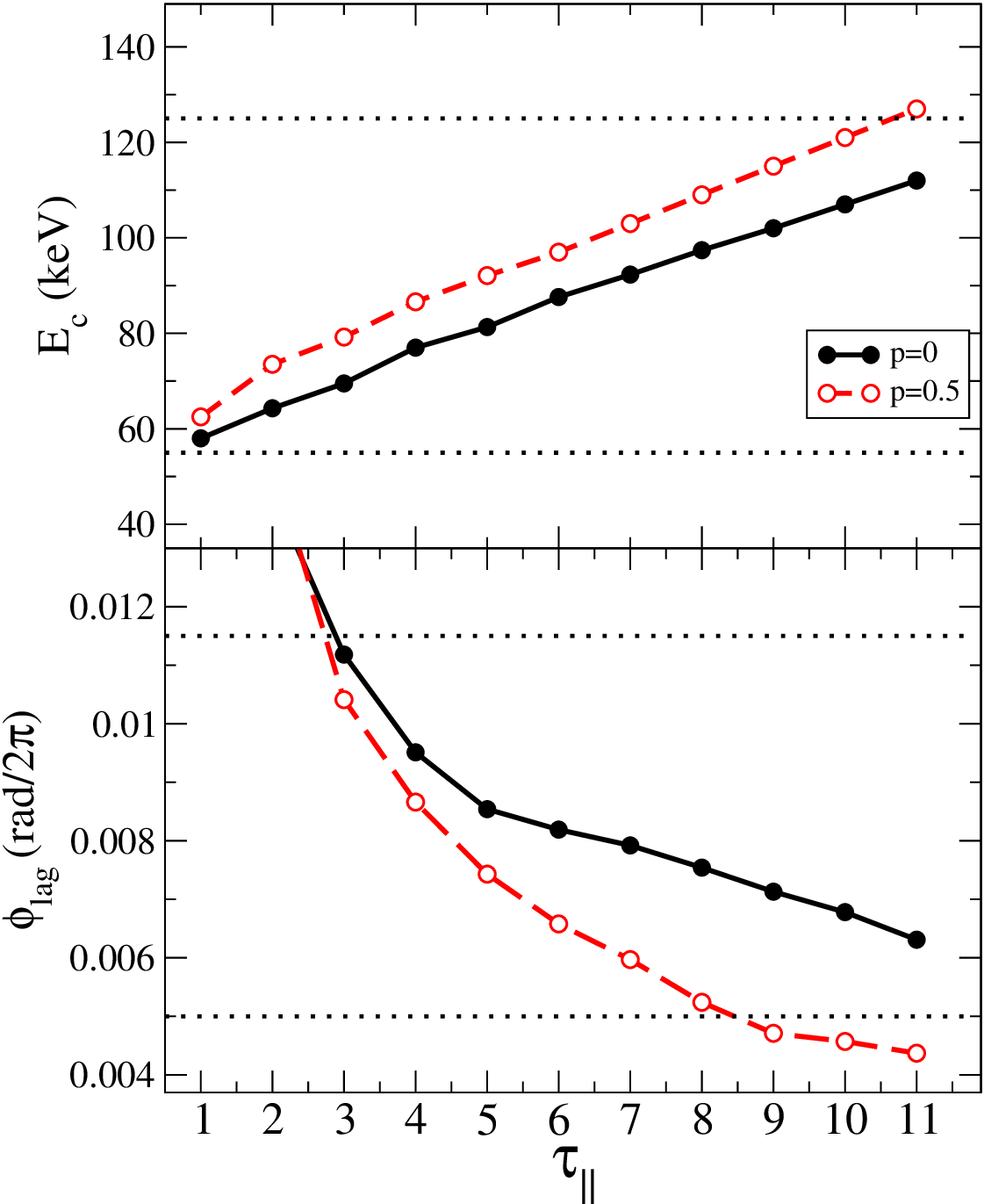}
   \caption{Cutoff energy and phase lag as functions of
   $\tau_{\parallel}$ when no acceleration zone (filled circles) 
   and when an acceleration zone is included (empty circles). 
   Each point represents a calculation with the same 
   $\gamma_{\rm min}=2.28$ ($v_{\perp}=0.41$) and 
   different $\tau_{\parallel}$.  The horizontal 
   dotted lines bracket the ranges of observed values as in \citet{motta09} 
   for $E_c$ and \citet{altamirano15} for  $\phi_{\rm lag}$.}
   \label{tauonly}
 \end{figure}

 \begin{figure}[!t]
   \centering
   \includegraphics[width=8cm]{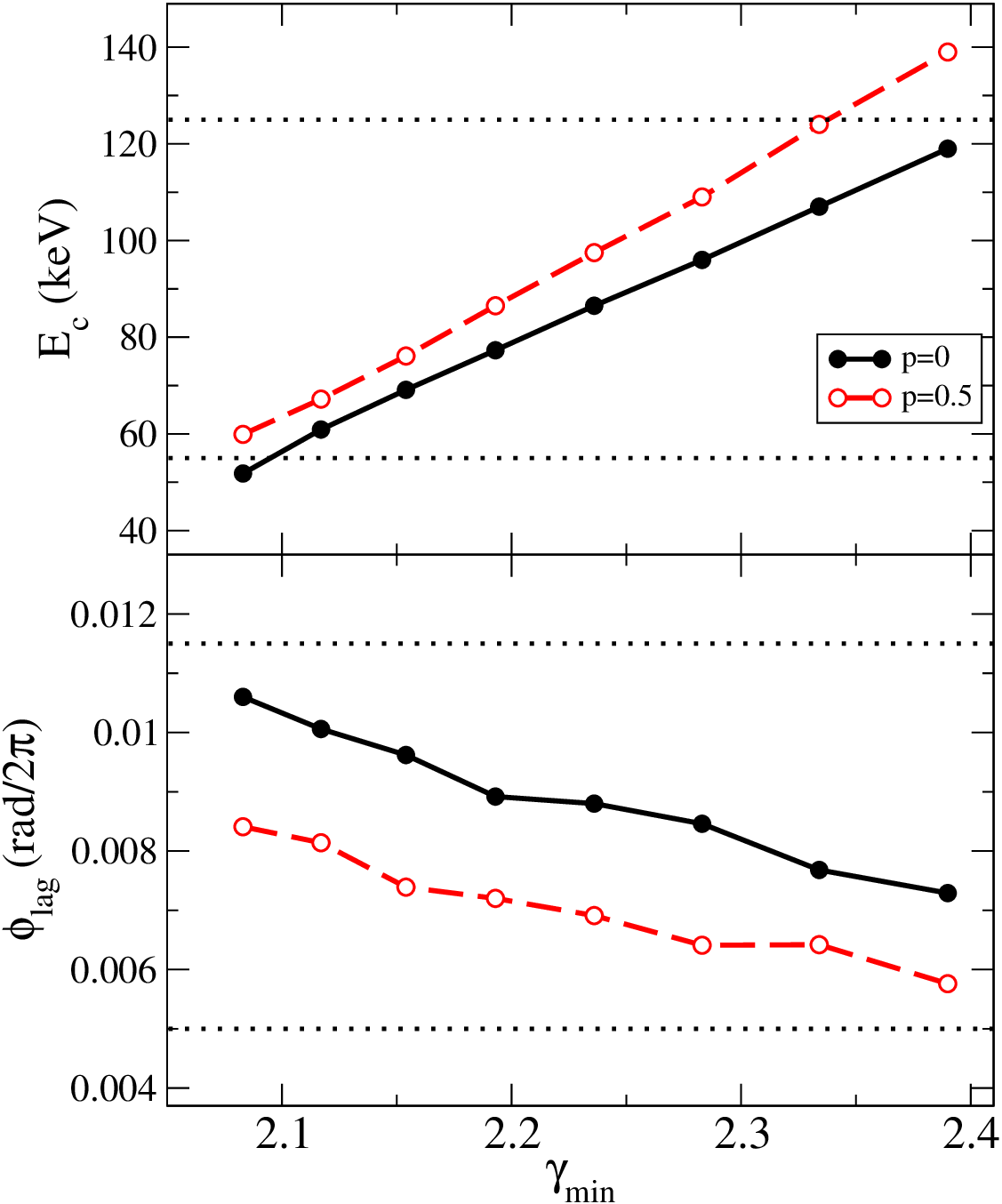}
   \caption{Cutoff energy and phase lag as functions of
   $\gamma_{\rm min}$ when no acceleration zone (filled circles) 
   and when an acceleration zone is included (empty circles). Each point
   represents a calculation with the same $\tau_{\parallel}=5$ and different
   $\gamma_{\rm min}$ ($v_{\perp}$). 
   The horizontal 
   dotted lines bracket the ranges of observed values as in \citet{motta09}  
   for $E_c$ and \citet{altamirano15} for $\phi_{\rm lag}$. }
   \label{velonly}
 \end{figure}

\section{Results}

The output of our Monte Carlo code consists of energy spectra and light
curves. In this work, the light curves were obtained for the energy bands
$2 - 5.7$ keV and $5.7 - 15$ keV to match those of \citet{altamirano15} for
the BHB GX 339--4.

The output energy spectrum has been fitted by a power-law with an
exponential cutoff, to extract the photon index $\Gamma$
and the cutoff energy $E_c$. Fourier analysis has been applied to the light
curves to obtain the phase lag of the hard X-rays with respect to the
soft ones as a function of Fourier frequency. The final phase lag
considered, $\phi_{\rm lag}$, is the average over the frequency range $0.01
- 5$ Hz.

Despite the fact that, in reality, the parameters $\gamma_{\rm min}$  and
$\tau_\parallel$ decrease simultaneously, for clarity of presentation,  we
will first vary $\tau_\parallel$, then  $\gamma_{\rm min}$, and at the end
both. The result that we wish to reproduce in this work is the {\em
decrease} of the cut off energy and the {\em increase} of the phase lag of
GX 339--4 as the X-ray intensity  increases and the source traverses the
hard state. As explained above,  we expect $\tau_\parallel$ and 
$\gamma_{\rm min}$ to decrease with time. However, for plotting purposes, we
left these two parameters, the independent variables of the plots, to 
increase toward the right in the $X$-axis, as is normal practice. For this
reason $E_c$ {\em increases} and $\phi_{\rm lag}$ {\em decreases} with
$\tau_\parallel$ and $\gamma_{\rm min}$. 

\subsection{Variation  of $\tau_\parallel$}

In Fig.~\ref{tauonly}, we show the variations of  $E_c$ and $\phi_{\rm
lag}$  as functions of the Thomson optical depth $\tau_\parallel$. The
solid line represents the case where no acceleration zone is considered 
(model 1, $p=0$), while the dashed one represents the case when an
acceleration zone is taken into account  (model 2, $p=0.5$). The Lorentz
factor of the electrons is fixed at  $\gamma_{\rm min}=2.24$ ($v_\parallel
= 0.8 c$, $v_\perp = 0.395 c$) and the rest of the parameters at their
reference values.  In both panels, the horizontal dotted lines bracket
the observed values of $E_c$ and  $\phi_{\rm lag}$, as
determined by  \citet{motta09} and \citet{altamirano15}.

It is important to stress that the variation of $E_c$ and
$\phi_{\rm lag}$ has the correct trend for both model 1 
and model 2.  On the other hand, variation of only $\tau_\parallel$ 
is not enough to reproduce simultaneously the observed ranges of variation 
of $E_c$ and $\phi_{\rm lag}$ for the same range of variation
of $\tau_\parallel$. 

Whereas $E_c$ nicely covers the whole range of observed variation  when the
optical depth $\tau_\parallel$ decreases from 9 to 1,  the phase lag
$\phi_{\rm lag}$ is outside the range  of the observations at low optical
depths, and lies short at high  optical depth. The introduction of an
acceleration zone (model 2, $p=0.5$) makes  the $\phi_{\rm lag}$ trend 
somewhat steeper, but again the two parameters do not cover the  observed
ranges of allowed values for the same range of optical depth.  Similar
curves result for values of $p$ different from 0.5, as long as
$p \simless 1$.

As it was shown in Papers I--IV, keeping the rest of the parameters to
their reference values and decreasing the density $n_0$ (or equivalently 
the Thomson optical depth along the axis of the jet $\tau_{\parallel}$, see
eq.~(2) in Paper IV), makes the emergent spectra softer.  
A decrase in the optical depth $\tau_\parallel$ also causes
a decrease in the cutoff energy $E_c$,
due to the reduced number of scatterings.  The effect
on the time lag or the phase lag is, however, less obvious.  
At large optical depth, the soft input photons 
penetrate the base of the jet, on average, a skin optical
depth of unity.  There, they are scattered, but due to the large density,
the mean free path is small and the photons sample a small region at the
base of the jet before they escape.  As the optical depth (or 
equivalently the density) decreases, the mean free path increases and at
small optical depths the photons sample the entire volume of the jet.
Thus, the time lag or the phase lag increases.

In the top panel of Fig. 1, one sees that for model 2 ($p=0.5$) $E_c$
is consistently larger than the corresponding value for model 1 ($p=0$).  
This is due to the fact that the mean number of scatterings of the photons
in model 2 is larger than that in model 1, which in turn is due to the 
higher density of the electrons at the base of the jet in model 2 than 
in model 1.

\subsection{Variation  of $\gamma_{\rm min}$}

In Fig.~\ref{velonly} we show the variation of $E_c$ and $\phi_{\rm lag}$
as functions of the Lorentz factor $\gamma_{\rm min}$.  The optical depth
is fixed to $\tau_\parallel=5$  and the rest of the parameters to their
reference values. Again, the trend followed by the computed quantities
matches  that of the observations. 

It is natural to expect that $E_c$ would decrease as the energy of the 
electrons decreases.  The  increase of $\phi_{\rm lag}$ as $\gamma_{\rm
min}$ decreases  requires an explanation.

At large $\gamma_{\rm min}$, i.e. large $v_\perp$, there is a tendency
for the photons after scattering to have directions with polar angle
$\theta$ closer to 90 degrees than to zero degrees.  Thus, the
random walk of the photons occurs close to the base of the jet, and
the time lag or phase lag is relatively small,
because the size of the jet is small there.  As 
$\gamma_{\rm min}$ decreases, i.e. $v_\perp$ decreases, the photons
after scattering have the tendency to have directions closer to
$\theta = 0$ than to $\theta = 90$ degrees.  Thus, the input photons
are pushed in the flow direction and are forced to sample the entire
volume of the jet, which results in relatively larger time lag or
phase lag.

As in Fig.~\ref{tauonly}, only $E_c$ covers the entire range of observed
values (indicated by the horizontal  dotted lines). The phase lag lies
short in one or both of the extremes. The introduction of an acceleration
zone  (model 2, $p=0.5$)  changes the amplitude of the lags substantially.
This is because in the acceleration zone $v_\parallel < v_0$ and scattering
in directions with $\theta$ closer to 90 degrees is enhanced. Thus, the
phase lag becomes smaller than in the case of no acceleration zone.

\subsection{Variation of $\tau_\parallel$ and $\gamma_{\rm min}$}

\begin{figure}[t]
  \centering
    \includegraphics[width=8cm]{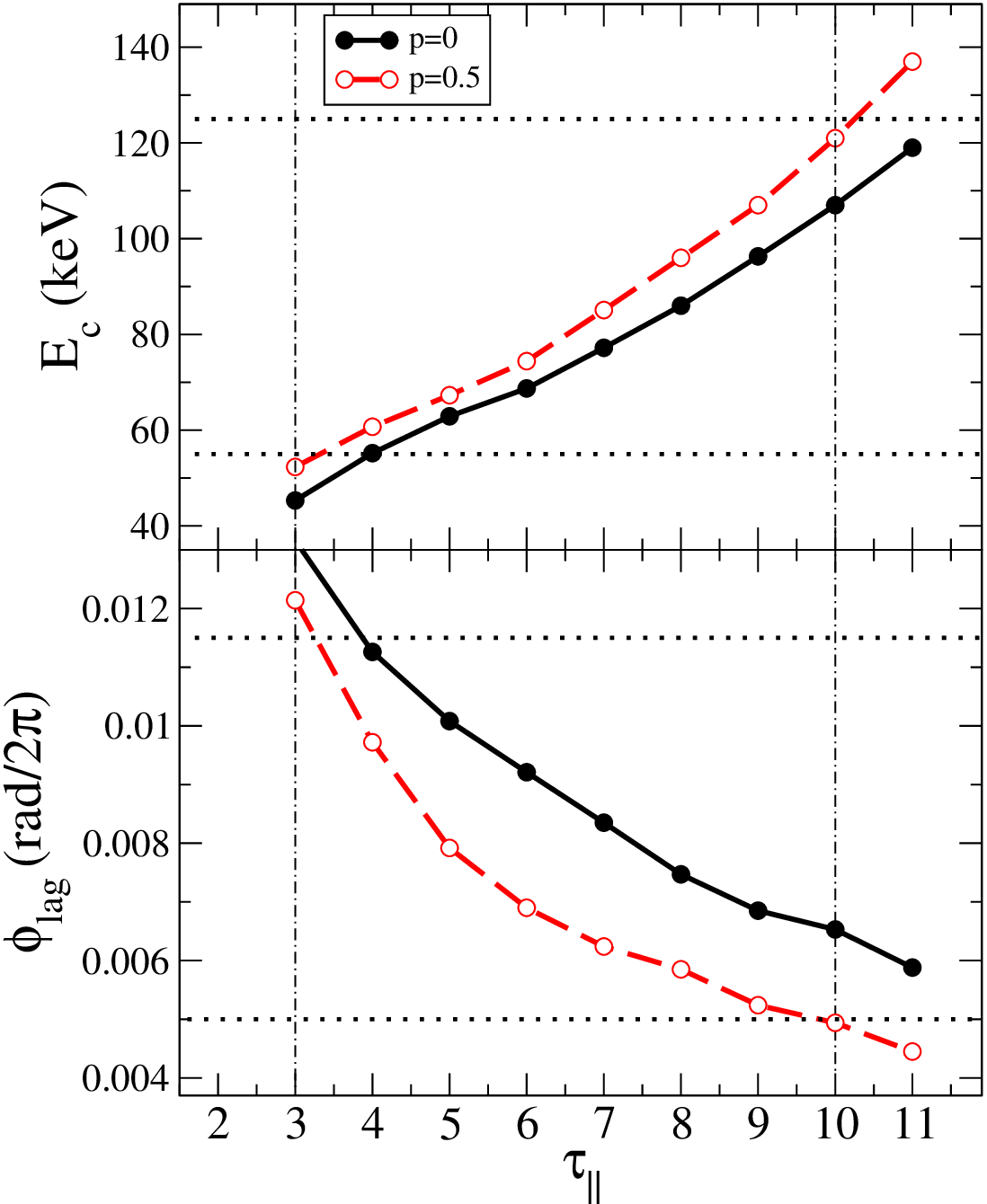}  
   \caption{Cutoff energy and phase lag as a function of 
   $\tau_\parallel$ when no acceleration zone (filled symbols)  and when an
   acceleration zone is included (open symbols). Each point represents a
   calculation with different 
   $\tau_{\parallel}$ and $\gamma_{\rm min}$. The horizontal  dotted lines
   bracket the ranges of observed values as in \citet{motta09}  for $E_c$
   and \citet{altamirano15} for $\phi_{\rm lag}$. The vertical dot-dashed
   lines show the range in $\tau_{\parallel}$ (hence in $\gamma_{\rm min}$
   also, see text) where the two quantities cover {\em simultaneously}
   the entire observed ranges. }
  \label{fig3}
 \end{figure}

Since, in reality, both $\tau_\parallel$ and $\gamma_{\rm min}$ decrease
simultaneously as the source moves in the hard  state, we varied them both
in the simplest way possible, one proportional to the other. In
Fig.~\ref{fig3}, we show  $E_c$ and $\phi_{\rm lag}$ as functions of
$\tau_\parallel$. In this Figure, $\gamma_{\rm min}$ is linked to
$\tau_\parallel$ by  $\gamma_{\rm min}=2.023 - 0.019 \,\,
\tau_\parallel$.


With two parameters, $\tau_\parallel$ and $\gamma_{\rm min}$, working in
the same direction, it is  not surprising that  we have now been able to
demonstrate that for $3 \le \tau_\parallel \le 10$, both computed
quantities  with model 2,  cover simultaneously the entire
observed ranges (see the vertical dot-dashed lines in Fig.~\ref{fig3}).

\begin{figure}[t]
  \centering
    \includegraphics[width=8cm]{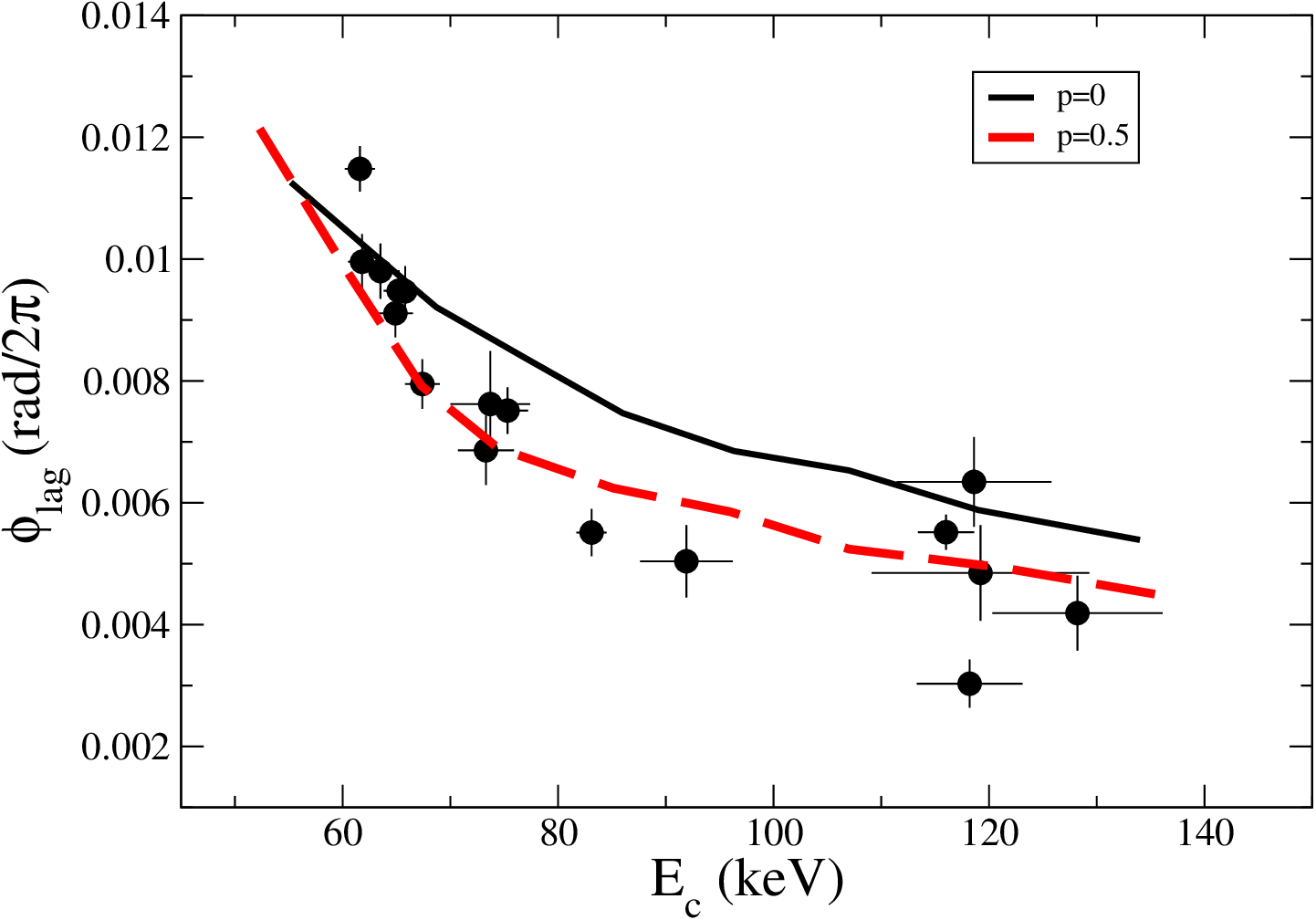}  
   \caption{Phase lags as a function of cutoff energy. Circles represent
   data from the observations. The cutoff energy
   and phase lag values were taken from \citet{motta09} and \citet{altamirano15},
   respecively. Solid and dashed lines correspond to the models shown in
   Fig.~\ref{fig3}.
    }
  \label{fig4}
 \end{figure}

To better illustrate the good agreement between the observations and
our model, we have plotted the phase lag as a function of the cutoff
energy in Fig.~\ref{fig4}. The data points come from the works of 
\citet{motta09} and \citet{altamirano15}. The solid and dashed lines
corresponds to the models shown in Fig.~\ref{fig3}.

\section{Summary and conclusion}

Observations from space-based X-ray telescopes over the last decades
have provided a wealth of data which have led to a revolution in our
understanding of black-hole binaries. Numerous studies of these
observations have allowed the characterisation of their spectral and
temporal properties and the definition of source states. Correlations
between the spectral and timing parameters impose tight  observational
constraints that any model that seeks to explain the observations must
address. Our jet model has so far been able to quantitatively explain  a
number of results regarding the hard state of black-hole binaries (Papers
I--IV):  {\em i)} the emergent spectrum from radio to hard X-rays, {\em
ii)}  the time(phase)-lags as a function of Fourier frequency, {\em iii)}
the flattening of the power spectra at high frequencies with increasing
photon energy and the narrowing of the autocorrelation function, {\em iv)} 
the correlation  observed in Cyg X-1 between the photon
index and the average time lag.

The works of \citet{motta09} and \citet{altamirano15} provide one more
stringent constraint: the  cutoff energy of the power law and the phase lag
of hard photons with respect to soft ones vary in unison as the black-hole
binary GX 339--4 evolves through the hard state and moves up in the
$q$-diagram \citep[see Fig.~1 in][]{altamirano15}.   In this work, we show
that  we can reproduce this correlation with a minor modification of our
model, namely the introduction of an acceleration zone. 

The correlation between the cutoff energy and the phase lag suggests that
the most likely origin of the lag is the same region as where the hard
X-rays are formed \citep{altamirano15}. Because the spectrum of the
radiation at the energies where the cutoff is measured (tens of keV) cannot
be formed in the accretion disk, the $E_{\rm c}-\phi_{\rm lag}$ correlation
strongly suggests that the lags are due to Comptonization. In our model,
Comptonization takes place in the jet. Hence we conclude that the phase lag
must originate in the jet. Another result that supports the association
between lag and jet stems from the fact that, at least for Cyg X--1, the lag
drops as the source moves into the soft-intermediate state \citep[see Fig.
6 in][]{altamirano15}, that is when the radio emission quenches.

\begin{acknowledgements}

We thank the anonymous referee for useful comments and especially for the
suggestion of Fig.~\ref{fig4}. We acknowledge useful discussions with
Dimitrios Gannios. This research has been supported in part by the
"RoboPol" project, which is implemented under the "ARISTEIA" Action of the
"OPERATIONAL PROGRAM EDUCATION AND LIFELONG LEARNING" and is co-funded by
the European Social Fund (ESF) and National Resources.

\end{acknowledgements}

\bibliographystyle{aa}
\bibliography{../jet}

\end{document}